\documentclass[aps,groupedaddress,10pt,amsmath,amssymb,twocolumn]{revtex4}
\usepackage{epsfig}
\usepackage{xspace}
\usepackage{amssymb,amsmath}
\usepackage{graphicx}

\begin{document}

\title{Coupled cluster benchmarks of water monomers and dimers extracted from DFT liquid water: the
importance of monomer deformations}
\author{Biswajit Santra$^1$}
\author{Angelos Michaelides$^{1,2}$}
\email{angelos.michaelides@ucl.ac.uk}
\author{Matthias Scheffler$^1$}
\affiliation{$^1$Fritz-Haber-Institut der Max-Planck-Gesellschaft, Faradayweg 4-6, 14195 Berlin, Germany \\
$^2$London Centre for Nanotechnology
and Department of Chemistry, University College London, London WC1E
6BT, UK}
\begin{abstract}
To understand the performance of popular density-functional theory (DFT)
exchange-correlation (xc) functionals in simulations of liquid water,
water monomers and dimers were extracted from a PBE simulation of liquid water
and examined with coupled cluster with single and double excitations plus
a perturbative correction for connected triples [CCSD(T)].
CCSD(T) reveals that most of the dimers are unbound compared to two gas phase equilibrium
water monomers, largely because monomers within the
liquid have distorted geometries.
Of the three xc functionals tested, PBE and BLYP systematically underestimate the
cost of the monomer deformations and consequently
predict too large dissociation energies between monomers within the dimers.
This is in marked contrast to how these functionals perform for an equilibrium
water dimer and other small water clusters in the gas phase, which only have moderately deformed monomers.
PBE0 reproduces the CCSD(T) monomer deformation energies very well and consequently the
dimer dissociation energies much more accurately than PBE and BLYP.
Although this study is limited to water monomers and dimers, the results reported
here may provide an explanation
for the overstructured radial distribution functions routinely observed in BLYP and PBE
simulations of liquid water
and are of relevance to water in other phases and to
other associated molecular liquids.
\end{abstract}

\maketitle

\vspace{2cm}
\textbf{I. Introduction}\\

Density-functional theory (DFT) has been widely used to study liquid water.
However, how well DFT with popular
exchange-correlation (xc) functionals such as PBE \cite{PBE} and BLYP \cite{Becke88,LYP} performs in describing the
structural and dynamic properties of liquid water is a matter of more than a little contention.
The debates, which are numerous, have hinged on issues such as the radial distribution functions (RDFs)
(in particular the O-O and O-H RDFs), diffusion coefficient, and average number of
hydrogen bonds (HBs).

It is now clear that most standard DFT molecular dynamics (MD) simulations
with PBE and BLYP predict an overstructured RDF compared to experiment.
By overstructured, we mainly mean that the first peak in the O-O RDF (referred to as $g_\mathrm{O-O}^{max}$) is higher than experiment.
Consequently the computed diffusion coefficient is too small and the average number of HBs too large.
Extended discussions on the magnitude and origin of the overstructuring can be found in e.g. Refs.
\cite{grossman_water_1,grossman_water_2,artacho,McGrath_05,McGrath_06,Mcgrath_jpcA_06,tuckerman_dvr_05,tuckerman_jcp_06,tuckerman_jcp_07,Todorova,
vandevondele_05,vandevondele_08,parrinello_09,sit_marzari_jcp_05,Lin_vdw_water_09,artacho_05,asthagiri_03,
am05_water_09,Kuo_04,sprik_jcp_96,silvestrlli_jcp_99,xantheas_JCP_09}.
In brief, some of the relevant factors include:
(i) The intrinsic error associated with a given xc functional (including an improper account
of van der Waals forces \cite{santra_08,Lin_vdw_water_09});
(ii) The omission of quantum nuclear effects \cite{parrinello_03,xantheas_06,voth_07,burnham_08,car_08};
and (iii) The simulation protocol, with relevant factors in this regard being:
(a) number of water molecules in the simulation cell \cite{parrinello_09};
(b) the density of the water within the cell \cite{McGrath_05,McGrath_06,Mcgrath_jpcA_06};
(c) basis set \cite{tuckerman_dvr_05,mattson_08};
(d) fictitious electron mass in Car-Parrinello MD simulations \cite{cpmd_prl_85};
and so on.
Since the first DFT MD simulation of liquid water in 1993 \cite{DFT_MD_93},
important strides have been made to understand how each
of the above factors impact upon the computed properties of liquid water.
However, simultaneously addressing all
issues that could account for the difference
between the experimental and theoretical RDFs and diffusion
coefficients is not practicable, not to mention the uncertainties that are present in the
experimental data itself \cite{soper_2000,T-Head-Gordon_06,soper_2007}.
Therefore, it has become common to attempt to shed light on the performance of
DFT xc functionals for treating water by investigating well defined
gas phase water clusters for which precise comparison
can be made to high level quantum chemistry methods.
This approach has been useful and allowed the intrinsic accuracy of many xc functionals to be precisely established
\cite{santra_08,sms,Truhlar_hexamer,BenchmarkNonbonded,X3LYP_dimer,X3LYP_hexamer_04,
joel_pbe,Tsuzuki,Novoa,xantheas_94,Tschumper_JCPA_06,shields_kirschner,Perdew_JCPB_05,truhlar_pbe1w,
truhlar_many-body_06,kim_jordan_94},
information that may be of relevance to liquid water.

With few exceptions \cite{Tschumper_JCPA_06,shields_kirschner,Perdew_JCPB_05,truhlar_pbe1w,truhlar_many-body_06},
previous gas phase benchmark studies of water clusters have focussed on exploring equilibrium or other
stationary point configurations of the gas phase intermolecular potential energy surfaces.
However, in the liquid the structures of water clusters and even the water monomers themselves can be considerably different from those
of gas phase clusters.
For example, the distribution of intramolecular O-H bond lengths in the liquid ranges from
$\sim$0.75 to $\sim$1.25 \AA\ \cite{soper_2000,soper_2007}.
Yet in gas phase water clusters
such as dimers to hexamers O-H bond lengths deviate by $<$0.05 \AA\  from the equilibrium water monomer
O-H bond length of 0.96 \AA\ \cite{sms,santra_08}.
Whether or not the performance of DFT xc functionals obtained from gas phase studies
on water clusters holds for the `deformed' structures
present in the liquid (referred to throughout this article as `deformed') remains an important open question.
Indeed there is already evidence that the benchmark reference data obtained from gas phase clusters does
not easily translate to the liquid.
For example, BLYP predicts a dissociation energy for the equilibrium gas phase water dimer that is 35 meV too small, yet
at a water density of 1 g/cm$^3$ it predicts a $g_\mathrm{O-O}^{max}$ that is about 5\%-15\% too high.
Similarly, PBE predicts the dimer dissociation energy to within 10 meV precision, yet yields an even
greater $g_\mathrm{O-O}^{max}$ than BLYP.
Related to this, MD simulations of liquid water have shown that the computed $g_\mathrm{O-O}^{max}$ can be considerably reduced
if the O-H bonds in the water monomers in the liquid are held rigid at some predefined bond length.
Specifically,  Allesch \emph{et al.} found that the PBE $g_\mathrm{O-O}^{max}$ decreased
by $\sim$10\%
upon going from fully relaxed
water monomers to a liquid with monomer O-H bonds fixed at $\sim$1 \AA~ \cite{galli_rigid_water}.
Likewise, Leung \emph{et al.} have shown through a careful and systematic series
of simulations that the length of the O-H bonds
for rigid water MD simulations directly correlates with $g_\mathrm{O-O}^{max}$ \cite{susan_06}:
as the intramolecular O-H bonds are allowed to lengthen, $g_\mathrm{O-O}^{max}$ increases.

The studies with rigid water and the realization that water monomer and cluster structures in the
liquid are likely to differ considerably from gas phase water clusters has prompted us to assess the
performance of DFT xc functionals on water structures more representative of those present in the
liquid.
To this end we report herein on the accuracy of three DFT xc functionals for various
deformed monomers and dimers taken from a PBE simulation of liquid water.
Two of the most popular generalized gradient approximation (GGA) xc functionals
for liquid water simulations (PBE and BLYP)
and one of the most accurate hybrid functionals
for small water clusters (PBE0 \cite{PBE0}) are assessed here.
As a reference, coupled cluster with single and double excitations plus
a perturbative correction for connected triples [CCSD(T)] is used with energies
extrapolated to the complete basis set limit (CBS).
The CCSD(T) reference calculations reveal that 75\% of the dimers extracted from within the first coordination shell
of the liquid are unbound relative to two equilibrium (gas phase)
water monomers.
This is mainly due to the large deformation of the monomers inside the liquid
compared to the gas phase equilibrium monomer structure.
PBE and BLYP consistently underestimate the cost of the monomer deformation,
specifically, O-H bond stretching.
As a consequence, both PBE and BLYP systematically overbind the deformed dimers
extracted from the liquid, by as much as 80 and 43 meV, respectively.
These errors are much larger
than the usual errors associated with these xc functionals for the gas phase equilibrium dimer \cite{sms}.
In general, the performance of PBE0 is superior to the two GGAs but noticeable errors are identified
for all functionals including PBE0 for the particularly long O-H bonds encountered at the shortest O-O separations.
Although this study is restricted to monomers and dimers (and in a sense resembles a highly limited cluster expansion
study of the liquid), the results reported here provide a possible explanation
for the overstructured RDFs routinely observed in BLYP and PBE simulations of liquid water.
The significance of these results to water in other phases and to
other associated molecular liquids is also briefly discussed.\\

\begin{figure}
 \begin{center}
     \includegraphics[width=8cm]{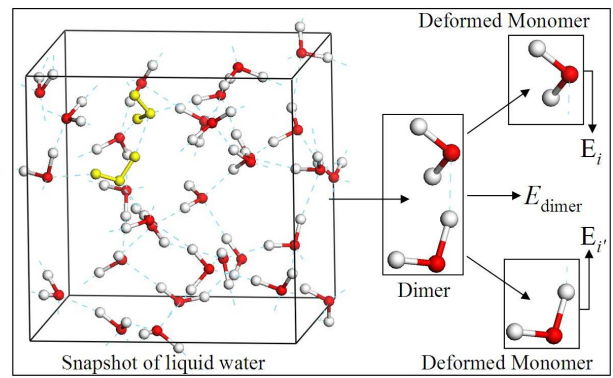}
 \caption{\label{fig1} (Color online) From a PBE MD simulation of liquid water, water dimers
 are extracted (e.g. the highlighted dimer in yellow).
 Single point energy calculations are then performed with CCSD(T), PBE, BLYP, and PBE0
 on the deformed dimers ($E_{\mathrm{dimer}}$)
 and the constituent deformed monomers ($E_i$, $E_{i^\prime}$). These energies are then used to
 evaluate the electronic dissociation energy of the dimers (Eqn. (4)) and the associated 1--body (Eqn. (2))
 and 2--body (Eqn. (3)) energies. The deformation of the monomers compared to a gas phase equilibrium monomer is also
 quantified with (Eqn. (1)).}
 \end{center}
\end{figure}

\textbf{II. Methods, procedures, and definition of parameters}\\

Several methods have been employed to study the water clusters examined here. We
now briefly describe some of the relevant computational details,
how the water monomers
and dimers are selected from the liquid, and then define the energetic and structural
parameters used in the subsequent analysis.\\

\textbf{A. Liquid water}\\

To generate water monomer and dimer structures representative of
those present in liquid water
a Born-Oppenheimer molecular dynamics
simulation of 32 D$_2$O molecules in a
periodic cubic box of length 9.8528 \AA\ was performed with the CPMD code \cite{cpmd}.
The PBE xc functional
was used along with hard pseudopotentials of Goedecker \emph{et al.} \cite{goedecker_pp}
and an associated plane wave energy cut-off of 125 Ry.
This simulation was run for 30 ps with an integration time step of 0.5 fs.
A Nos\'e-Hoover chain thermostat was used to maintain a target temperature of 330 K.

\begin{figure}
 \begin{center}
     \includegraphics[width=8cm]{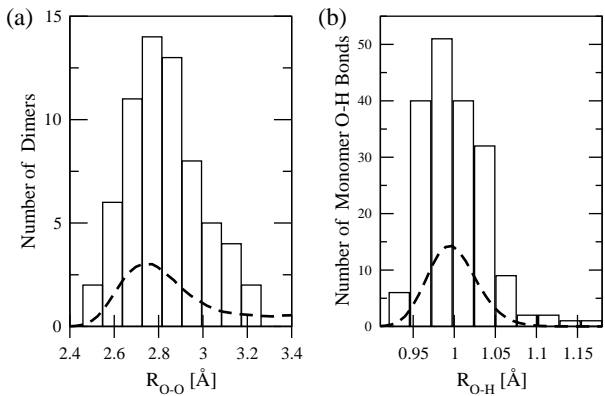}
 \caption{\label{fig2} (a) Distribution of the number of water dimers selected from the PBE MD
 simulation of liquid water (plotted as bars) as a function of the O-O separations ($R_{\mathrm\mathrm{O-O}}$)
 within the dimer.
 (b) Distribution of the number of O-H bonds for the water monomers selected from the liquid  (plotted as bars).
 The dashed lines represent the corresponding O-O and O-H RDFs from the same PBE
 liquid water simulation. Only dimers within the first coordination shell of liquid water are considered
 here.}
 \end{center}
\end{figure}

Water monomers and dimers were then
extracted from the MD simulation.
To get an uncorrelated sample of structures, 6-7 dimers were selected
each 2 ps over the last 20 ps of the MD trajectory.
In total 66 bonded dimers were selected (comprising 92 individual monomers).
The criteria we followed for selecting dimers were that: (i) they were from within the first coordination
shell of the O-O RDF,
i.e., all chosen O-O distances ($R_{\mathrm\mathrm{O-O}}$) are
$\le$3.4 \AA;
and (ii)
the distribution of all 66 $R_{\mathrm\mathrm{O-O}}$ of the dimers resembles the O-O RDF
for the first coordination shell.
Fig. \ref{fig2}(a) illustrates that the distribution of dimers selected
does indeed resemble the computed O-O RDF of our MD simulation reasonably well.
%
As an independent check we note that the distribution in the values of the intra--molecular O-H
bond lengths associated with all the selected water molecules is also in reasonably
good agreement with the
first peak of our computed O-H RDF of liquid water [Fig. \ref{fig2}(b)].\\

\textbf{B. Clusters}\\

The PBE water monomers and dimers extracted from the liquid water simulation were
then examined with a few DFT xc functionals and CCSD(T).
Throughout this study structures collected from the liquid water simulations are used for single point
energy calculations with the various methods and
no geometries are optimized, unless explicitly stated otherwise.
The DFT calculations on the gas phase water monomers and dimers were performed
with the GAUSSIAN03 code \cite{g03}, using large Dunning correlation consistent
aug-cc-pV5Z basis sets \cite{dunning_aug-cc-pv5z}.
We have shown before that for DFT xc functionals such as the ones considered
here, this basis set is large enough to get dissociation energies within about 1 meV/H$_2$O
of the CBS limit \cite{sms}.
Results from just three xc functionals will be reported, specifically two GGAs that
are widely employed in DFT simulations of liquid water (PBE and BLYP), and
the hybrid PBE0 functional, which is one of the most accurate functionals in predicting the
absolute dissociation energies of small gas phase water clusters (dimer to hexamer) \cite{sms,santra_08}.

The CCSD(T) calculations were performed with the  NWChem code \cite{nwchem}
with localized Gaussian basis sets.
Specifically,
aug-cc-pVnZ basis sets (n = T, Q, and 5) were
used and the resultant energies extrapolated to the complete basis set limit (CBS)
with the same standard heuristic schemes as employed by us before \cite{sms,santra_08}.
CCSD(T)/CBS is the theoretical `gold standard' for systems of the size considered here and, in the
following, differences between a given xc functional and CCSD(T) are referred to as errors
with that xc functional.
In total $>$600 CCSD(T) calculations have been performed for the reference data presented in this paper.
\\


\textbf{C. Definition of Parameters}\\

In order to quantitatively compare the structure of the molecules extracted from the liquid
to a gas phase equilibrium monomer we define a quantity $S_d$, the deformation, as,
\begin{equation}
 S_d = \sqrt{\sum_N (\textbf{R}_N-\textbf{r}_N)^2} \quad ,
\label{eqn_Sd}
\end{equation}
where $N$ is the number of atoms, $\textbf{R}$ and $\textbf{r}$ denote the coordinate vectors of deformed
and gas phase equilibrium monomer structures, respectively.
%

Several energy terms will appear repeatedly and it is also useful to define them here.
The one-body energy ($E_{1b}$) of a water monomer
is calculated as,
\begin{equation}
 E_{1b} = E_i - E_{\mathrm{equilibrium}} \quad ,
\label{eqn_1b_energy}
\end{equation}
where $E_{\mathrm{equilibrium}}$ is the energy of the gas phase water monomer at equilibrium
and $E_i$ is the energy of a deformed monomer.
The two-body energy ($E_{2b}$) of a dimer is defined as:
\begin{equation}
 E_{2b} = E_{\mathrm{dimer}} - E_i - E_{i^\prime} \quad ,
\label{eqn_2b_energy}
\end{equation}
where $E_{\mathrm{dimer}}$ is the total energy of the dimer.
The electronic dissociation energy ($D_{e}$) of the dimers is given by,
\begin{equation}
 D_{e} = E_{\mathrm{dimer}} - 2 \times E_{\mathrm{equilibrium}} \quad .
\label{eqn_disso_energy}
\end{equation}
Fig. \ref{fig1} schematically illustrates each of the above energetic quantities
and the overall procedure used in this study.
Since the structures considered here have been taken from a PBE liquid water simulation,
$E_{\mathrm{equilibrium}}$ is calculated
with a PBE structure \cite{note_PBE_monomer}.
The error conceded by DFT xc functionals ($\Delta E$) in comparison to CCSD(T)/CBS is given as,
\begin{equation}
\Delta E = E_{\mathrm{CCSD(T)}} - E_{\mathrm{DFT}} \quad ,
\label{eqn_error}
\end{equation}
where $E_{\mathrm{CCSD(T)}}$ and $E_{\mathrm{DFT}}$ are energies obtained from CCSD(T)/CBS and DFT,
respectively.\\

\textbf{III. Results}\\

Now we examine the water monomers extracted from the liquid, focusing on the cost to go from the gas phase equilibrium
monomer structure to the deformed structures present in the liquid. Following this we consider the
water dimers.
In each case we compare the results of the various DFT xc functionals to the CCSD(T)/CBS references.\\

\begin{figure}
 \begin{center}
     \includegraphics[width=8cm]{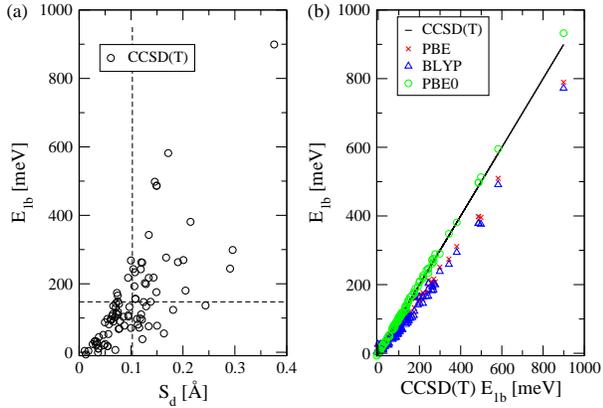}
 \caption{\label{fig3} (Color online) (a) CCSD(T) one-body energy ($E_{1b}$) versus deformation ($S_d$)
 for water monomers taken from PBE liquid water.
The horizontal and vertical dashed lines indicate the mean value of $E_{1b}$ and the mean value of the deformation,
respectively. (b) Comparison of $E_{1b}$ of PBE, BLYP, and PBE0 with CCSD(T).}
 \end{center}
\end{figure}

\textbf{A. Monomers}\\

To begin, CCSD(T) was used to establish the relative energies of the monomers taken from the
liquid compared to the gas phase equilibrium monomer, i.e., CCSD(T) $E_{1b}$ energies
were computed for all 92 monomers.
As can be seen from Fig. \ref{fig3}(a) these are distributed in a very large range from $\sim$0 to +900 meV
with a mean value of +147 meV.
Thus on average the monomers extracted from the liquid are 147 meV less stable than the gas phase equilibrium monomer,
a surprisingly large energy.
In quantifying the amount of deformation (Eqn. \ref{eqn_Sd}) for each monomer we find, as expected, a
general increase in $E_{1b}$
with the extent of deformation [Fig. \ref{fig3}(a)].
The average deformation of the monomers is $\sim$0.1 \AA, which gives us a measure of how deformed
water monomers are in a PBE liquid water structure.

\begin{figure}
 \begin{center}
     \includegraphics[width=8cm]{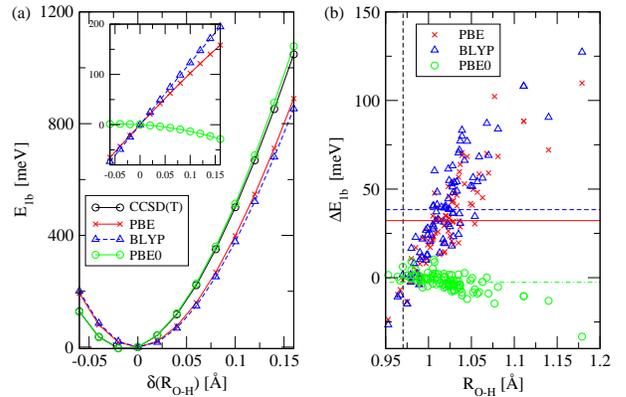}
 \caption{\label{fig4} (Color online)
 (a) Variation in the one-body energy ($E_{1b}$) with
 the O-H bond lengths
 of a water monomer calculated with CCSD(T), PBE, BLYP, and PBE0.
 The inset shows the differences in $E_{1b}$ of the three
 xc functionals compared to CCSD(T). (b) Errors in $E_{1b}$ ($\Delta E_{1b}$) for the deformed monomers selected from liquid
 water as a function of the longest O-H bond of each monomer.
 The vertical dashed line indicates the gas phase equilibrium O-H bond length (0.97 \AA) of a monomer (optimized with PBE) and
 the horizontal solid, dashed, and dash-dotted lines represent the average errors of
 PBE, BLYP, and PBE0, respectively.
 Here, a positive error of $E_{1b}$ indicates that it is too easy to stretch O-H bonds of the monomers with
 a given xc functional compared to CCSD(T).}
 \end{center}
\end{figure}

Now we consider how the deformation energies computed with PBE, BLYP, and PBE0 compare to CCSD(T).
This is shown in Fig. \ref{fig3}(b), a parity plot of $E_{1b}$ for the three xc functionals
compared to CCSD(T).
Immediately it can be seen that
the performance of the GGAs is markedly different from the hybrid PBE0 functional.
Specifically, for PBE and BLYP, $E_{1b}$ is systematically too small compared to CCSD(T).
On average the PBE and BLYP deformation energies are
32 and 38 meV, respectively,
smaller than that obtained from CCSD(T).
The size of the error simply increases with the
total CCSD(T) $E_{1b}$ [Fig. \ref{fig3}(b)] and for the largest deformations is on the order of 100 meV.
Remembering that the monomer deformation is, of course, an endothermic process, smaller values of $E_{1b}$
therefore indicate that
\emph{it is too easy to deform monomers to their liquid structures with PBE or BLYP compared
to CCSD(T)}.
In contrast to the GGAs, PBE0 produces $E_{1b}$ in excellent agreement with CCSD(T) with a mean error
of only
--3 meV.
This small negative error reveals that it is marginally too expensive to deform the monomers to
their liquid water structure with PBE0 compared to CCSD(T).
Since the only difference between PBE and PBE0 is the 25\% Hartree-Fock (HF) exchange in the latter,
we conclude that
the inclusion of exact exchange remedies the large error in $E_{1b}$ almost completely. Why this is so will
be discussed in section \textbf{IV}.

\begin{table}
\caption {\label{tb1} Computed harmonic vibrational frequencies for a water monomer.
$\nu_1$ and $\nu_2$ are the asymmetric and symmetric O-H stretching modes
and $\nu_3$ is the H-O-H bending mode.
All values are in cm$^{-1}$ and calculated with an aug-cc-pVTZ basis set.}
\begin{ruledtabular}
\begin{tabular}{cccc}
    & $\nu_1$ & $\nu_2$ & $\nu_3$ \\
\hline
CCSD(T)  & 3921 & 3812 & 1648 \\ 
PBE      & 3800 & 3696 & 1593  \\
BLYP     & 3756 & 3655 & 1596  \\
PBE0     & 3962 & 3856 & 1633  \\
\end{tabular}
\end{ruledtabular}
\end{table}

The monomers extracted from the liquid have both modified bond lengths and H-O-H internal angles.
In order to understand in detail
where the errors in $E_{1b}$ for the GGAs come from
we carried out a simple series of tests where bond lengths and the internal angle were varied independently.
The tests show that the main error in the GGAs comes from the bond stretching.
For example Fig. \ref{fig4}(a) shows that the symmetric stretching of the O-H bonds of a water monomer costs
much less energy with PBE and BLYP compared to CCSD(T).
The errors increase almost linearly with the stretching [inset Fig. \ref{fig4}(a)] and are as large
as $\sim$200 meV when the O-H bonds are 0.16 \AA\ longer than the gas phase equilibrium
bond length of 0.97 \AA.
A bond stretch of 0.16 \AA\ may sound like a lot but monomers with O-H bonds even as
long as 1.18 \AA\ are present in our PBE MD simulation and in experiment and in \emph{ab initio} path integral simulations
even longer O-H bonds are observed \cite{soper_2007,car_08}.
As we saw for the structures taken from the liquid, PBE0 is in very good agreement with CCSD(T) and
even for the longest O-H bond of 1.18 \AA\ comes within 34 meV of CCSD(T).
In addition alterations of the H-O-H angle were considered but this makes much less of
a contribution to the error in the
xc functionals than what we find for bond stretching.
For example, increasing
(decreasing) the bond angle by 15$^{\circ}$ causes a maximum error of 15 meV (--8 meV)
with BLYP and even smaller errors for the two other functionals.

The above tests establish that an inaccurate description of bond stretching is the main origin of the
error in $E_{1b}$ for the GGAs.
Returning to the structures taken from the liquid we therefore
plot in Fig. \ref{fig4}(b) the $E_{1b}$ error against the length of the longest O-H bond for each monomer.
As with the systematic deformations of the equilibrium monomer, the errors in $E_{1b}$ increase
almost linearly with the O-H bond length for the GGAs and for PBE0 they remain
very close to zero except at the longest distances.
Thus it can be inferred that monomers inside liquid water are energetically
too easy to stretch for both PBE and BLYP.
We note that a careful series of tests taking water clusters from a BLYP MD
simulation along with subsequent tests with CCSD(T)
established that none of the conclusions arrived at here are altered if BLYP structures are
used \cite{note_blyp_dimers}.

Before moving on to the dimers, we point out that the discrepancies established here between
the three xc functionals and CCSD(T) correlate well with the errors in the computed harmonic vibrational
frequencies of an isolated water monomer (Table \ref{tb1}).
Specifically for the two stretching frequencies PBE and BLYP are $\sim$115 to
$\sim$160 cm$^{-1}$ ($\sim$3-4\%) softer than CCSD(T) (Table \ref{tb1}), whereas,
PBE0 is only  $\sim$45 cm$^{-1}$  ($\sim$1\%) harder.
This one to one correspondence between error in harmonic vibrational frequencies and $E_{1b}$
may also hold for other xc functionals and may therefore provide a cheap diagnostic to estimate in advance how
reliably an xc functional will be at the determination of $E_{1b}$.\\

\begin{figure}
     \begin{center}
         \includegraphics[width=8cm]{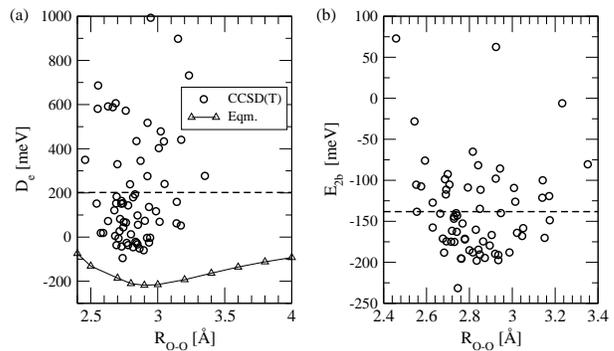}
     \caption{\label{fig5} (a) CCSD(T) total dissociation energies ($D_e$) of
     the water dimers taken from the PBE liquid water simulation and the dissociation energy curve for a fully optimized
     gas phase dimer (Eqm.) as a function of the O-O distance ($R_{\mathrm\mathrm{O-O}}$);
     (b) CCSD(T) two-body energies ($E_{2b}$) as a function
     of $R_{\mathrm\mathrm{O-O}}$ for the same dimers.
     The dashed lines represent the mean values of $D_e$ and $E_{2b}$ in panels (a) and (b), respectively.}
     \end{center}
\end{figure}

\begin{table*}
\caption {\label{tb2} Mean values of the one-body, two-body, and dissociation energies of
the deformed dimers (selected from the liquid) and the corresponding values for the gas phase equilibrium
dimer with CCSD(T), PBE, BLYP, and PBE0.
The differences between each DFT xc functional and CCSD(T)/CBS are given in parenthesis.
Energies are in meV.}
\begin{ruledtabular}
\begin{tabular}{cccc|ccc}
&\multicolumn{3}{c|}{deformed} & \multicolumn{3}{c}{Gas phase equilibrium}\\
\hline
    & D$_e$ ($\Delta$D$_e$) & E$_{1b}$ ($\Delta$E$_{1b})$ & E$_{2b}$ ($\Delta$E$_{2b}$)
    & D$_e$ ($\Delta$D$_e$) & E$_{1b}$ ($\Delta$E$_{1b})$ & E$_{2b}$ ($\Delta$E$_{2b}$) \\
\hline
CCSD(T) & 201.9        & 339.2        & --137.2        & --211.6          & 9.7       & --221.4 \\
PBE     & 121.9 (80.0) & 268.0 (71.2) & --146.1 (8.8)  & --219.9 (8.3)  & 3.7 (6.0) & --223.6 (2.2) \\
BLYP    & 159.3 (42.6) & 254.4 (84.8) & --95.1 (--42.1)& --178.7 (--32.9) & 2.4 (7.3) & --181.1 (--40.3) \\
PBE0    & 200.7 (1.2)  & 346.1 (-6.9) & --145.3 (8.1)  & --213.5 (1.9)  & 9.7 (0.0) & --223.2 (1.8)\\
\end{tabular}
\end{ruledtabular}
\end{table*}
%


\textbf{B. Dimers} \\

Now we move to the dimers extracted from the liquid, discussing first what CCSD(T) reveals about
the stability of the dimers and then considering how well the three functionals perform.
In Fig. \ref{fig5}(a) the CCSD(T) dissociation energies are plotted as a function of the O-O
distance within each dimer.
Also reported is the equilibrium (i.e., fully optimized) CCSD(T) dissociation energy curve for a gas phase water dimer.
As expected the equilibrium dimer binding energy curve provides a lower bound for the
dissociation energies of the deformed dimers, which at each
particular value of $R_{\mathrm\mathrm{O-O}}$  exhibit a range of values reflecting the
range of dimer structures in the liquid.
More importantly, Fig. \ref{fig5}(a) provides an overview of the range of
dissociation energies for water dimers found inside the first coordination
shell of PBE liquid water.
The range is large: from --95 to +993 meV, with the mean value being +201 meV.
Indeed 75\% of the dissociation energies are positive,
i.e., 75\% of the dimers are unbound compared to
two gas phase equilibrium water monomers.
Upon decomposing the dissociation energies into the one- and two-body contributions
we find that the average $E_{1b}$ is 339 meV
and the average $E_{2b}$ is --137 meV (Table \ref{tb2}).
Note that the average value of $E_{1b}$ for the dimers is, of course, about twice $E_{1b}$
for the monomers discussed above.
Also note that this is a considerably larger $E_{1b}$ than for the gas phase equilibrium water dimer,
which is only 10 meV (Table \ref{tb2}).
The two-body energy gives the binding between the water molecules and
97\% of the dimers have an attractive $E_{2b}$ [Fig. \ref{fig5}(b)].
The average value of $E_{2b}$ at
--137 meV
is somewhat smaller than the corresponding value for the gas phase equilibrium dimer of
--221 meV.
Since the total dissociation energy for the dimers is just the sum of $E_{1b}$ and $E_{2b}$,
it is quite obvious therefore that $E_{1b}$ plays the major role in destabilizing
the dimers.

\begin{figure}
 \begin{center}
     \includegraphics[width=8cm]{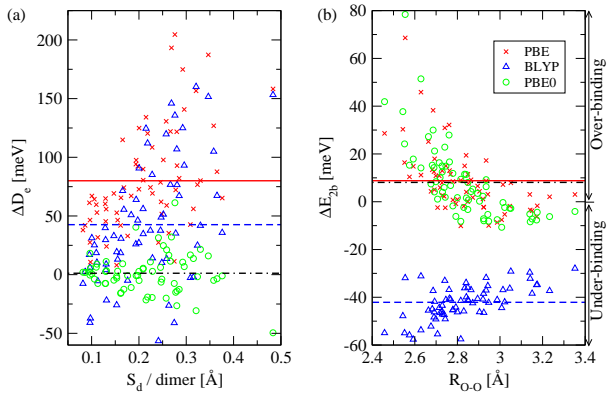}
 \caption{\label{fig6} (Color online) (a) Errors in dimer dissociation energy ($\Delta D_e$) as
 a function of the sum of the deformation of the two monomers within each dimer for PBE, BLYP, and PBE0.
 (b) Errors in the two-body energy ($\Delta E_{2b}$) from PBE, BLYP, and PBE0 as a function of
 the O-O distance ($R_{\mathrm\mathrm{O-O}}$) within the dimers. Horizontal solid, dashed, and
 dash-dotted lines represent the average errors of PBE, BLYP, and PBE0, respectively.}
 \end{center}
\end{figure}

Coming back to the performance of the xc functionals,
Fig. \ref{fig6}(a) reports the error in the dissociation energies for each dimer.
It can be seen that overall PBE0 performs very well, yielding an average error of
1 meV.
BLYP and PBE, on the other hand, yield quite large average errors of 43 and 80 meV, respectively.
This behavior differs significantly from how these two functionals perform for the gas phase equilibrium water dimer,
where errors of only --33 and 8 meV are obtained (Table \ref{tb2}).
Thus we arrive at a key result, the performance of the two GGA functionals for the
deformed dimer structures is inferior to what it is for the gas phase equilibrium dimer,
with both functionals substantially overbinding the dimers taken from the liquid.
Table \ref{tb2} reports the key quantities that allow us to understand these results and why they
contrast to the equilibrium gas phase dimers.
As one might anticipate from section \textbf{III A}, the key is the one-body deformation energy.
In the gas phase equilibrium dimer the absolute value of $E_{1b}$ is small (10 meV) and the resultant errors
even smaller (Table \ref{tb2}).
Thus the performance of a functional for the gas phase equilibrium dimer is dominated by $E_{2b}$,
which is accurately described with PBE and PBE0 (8 and 2 meV errors, respectively)
and underestimated by some 33 meV with BLYP.
However, as we have seen in the structures taken from the liquid, $E_{1b}$ is large [339 meV from CCSD(T)]
and the associated errors from the GGA xc functionals become significant.
Specifically, since both BLYP and PBE predict that the one-body deformation energy is much too small
and predict either too weak or about right two-body energies then the total dissociation energies
come out too large.
BLYP proves to be more accurate than PBE simply because of more favorable cancelations of errors in $E_{1b}$
and $E_{2b}$ (Table \ref{tb2}).
Since BLYP is generally considered to produce too weak HBs between
water molecules \cite{sms,Novoa,Tsuzuki,X3LYP_dimer}
the overbinding observed here is remarkable.
The obvious relevance of this finding to liquid water
will be discussed below.

\begin{figure}
 \begin{center}
     \includegraphics[width=8cm]{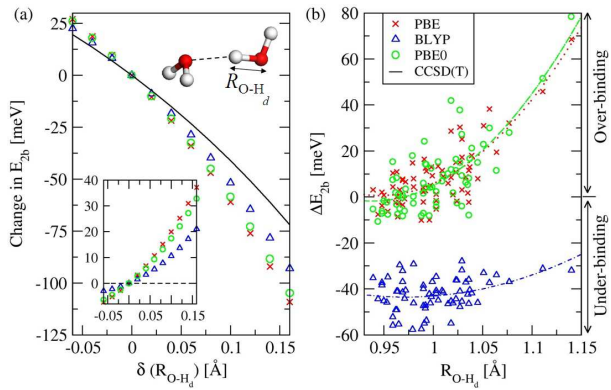}
 \caption{\label{fig7} (Color online) (a) Compared to the gas phase equilibrium dimer, the change in the two-body
 energy ($E_{2b}$) obtained from the systematic variation of the covalent
 O-H bond of the donor `H' atom [$\delta(R_{\mathrm{O-H}_d})$] keeping all other atoms fixed. The inset
 shows the difference between the DFT xc functionals and CCSD(T).
 (b) Error in the two-body ($E_{2b}$) energy from DFT compared to CCSD(T) as a function of the
 $R_{\mathrm{O-H}_d}$, obtained from the dimers from liquid water.
 Here positive values refer to a stronger two-body interaction.
 The dashed, dotted, and dash-dotted lines are quadratic fits to the PBE0, PBE, and BLYP data, respectively.}
 \end{center}
\end{figure}

A characteristic feature of HBs between water molecules is that the covalent
O-H bonds of the donor molecules ($R_{\mathrm{O-H}_d}$) are elongated \cite{Jeffrey_book}.
The elongation is a result of charge transfer from the acceptor
water molecule to the O-H $\sigma^*$ antibonding orbital of the donor
molecule \cite{hobza_chem_rev_00,M-head-gordon_09}.
Since we find that it is too easy to stretch an O-H bond with the GGAs, one can anticipate that this will
further influence the strength of the HBs formed and in particular $E_{2b}$.
To investigate this we return to the gas phase equilibrium water dimer as a test case
and systematically stretch the O-H$_d$ bond whilst keeping all other atoms fixed.
Fig. \ref{fig7}(a) plots the change in $E_{2b}$ as a function of the O-H$_d$ bond length
with CCSD(T) and the three xc functionals.
Clearly all methods predict that
as the O-H$_d$ bond increases so too does $E_{2b}$.
However, all three xc functionals predict too rapid an increase compared to CCSD(T).
This is best seen by the inset in Fig \ref{fig7}(a) which displays the error in the change of $E_{2b}$
as a function of $R_{\mathrm{O-H}_d}$ compared to CCSD(T).
Likewise, $E_{2b}$ increases slightly too rapidly with the three xc functionals for the dimers extracted
from the liquid [Fig. \ref{fig7}(b)]; this is particularly apparent for PBE and PBE0.
Thus in addition to it being too easy to stretch an O-H bond with BLYP and PBE,
for all three xc functionals the magnitude of the change in $E_{2b}$ upon stretching is
too great, further contributing to the overbinding of dimers with long
O-H$_d$ bonds. \\

\textbf{IV. Discussion} \\

\begin{figure}
 \begin{center}
     \includegraphics[width=5cm]{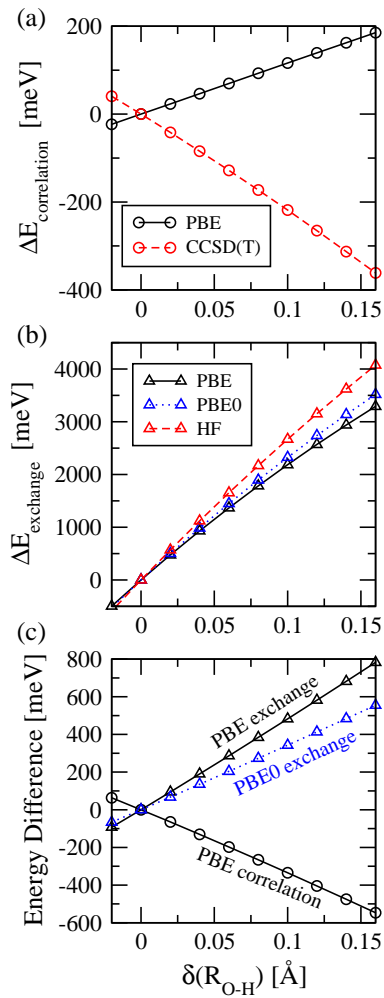}
 \caption{\label{fig8} (Color online) Variation in (a) correlation energy ($\Delta E_{\mathrm{correlation}}$)
 and (b) exchange energy ($\Delta E_{\mathrm{exchange}}$) with O-H bond length for a gas phase water molecule.
 (c) Variation in the difference of DFT correlation and
  exchange energies in comparison to the CCSD(T) correlation and HF exact exchange energies, respectively,
  as a function of the monomer O-H bond length. Positive and negative values refer to energy loss and gain, respectively.}
 \end{center}
\end{figure}

It is clear from the last section that, despite the $E_{2b}$ errors for the longest O-H$_d$ bonds, the overall
performance of PBE0 is superior to that of the GGAs.
To understand the origin of the difference between PBE and PBE0 the variation in
the exchange and correlation energies was examined upon going
from the gas phase equilibrium to the deformed water monomer structures (Table \ref{tb3}).
The variations in DFT exchange and correlation energies are then compared with the
full HF exact exchange and CCSD(T) correlation.
We note that the physical interpretation of exchange and correlation differs from DFT (PBE) to CCSD(T)
and so use the data reported in Table \ref{tb3} and Fig. \ref{fig8} merely in the hope of
obtaining some general qualitative insight.
The basic finding from CCSD(T) is that upon going from the gas phase equilibrium to the deformed
monomers there is a gain in the (negative) correlation energy
and a loss in exchange energy [Fig. \ref{fig8}(a),(b)].
Naturally, the absolute change in the exchange energy is far greater
than that in the correlation energy.
The two DFT xc functionals predict a loss in the exchange energy but in contrast to CCSD(T)
also a \emph{loss} in the correlation energy, i.e., there is less correlation with PBE in the
deformed monomers compared to the equilibrium monomer in the gas phase.
Thus, in terms of the correlation energy, PBE/PBE0 predicts qualitatively different
behavior from CCSD(T) upon bond stretching.
However, the missing correlation in the deformed monomers is compensated for
by differences from CCSD(T) in the exchange energy.
In PBE0, which predicts exchange energies in better agreement with HF (CCSD(T)) [Fig. \ref{fig8}(b)], the missing
correlation is compensated by missing exchange so that overall the total energy changes
are, as we have seen, very similar for PBE0 and CCSD(T).
In PBE, however, the lack of correlation is not sufficient to compensate for the larger
underprediction of the exchange energy [Fig. \ref{fig8}(b)].
Thus we find that PBE0 is superior to PBE simply because of a more favorable cancelation of the
differences of exchange and correlation from CCSD(T)  [Fig. \ref{fig8}(c)].

More generally the poor description of covalent O-H bond stretching observed here with PBE and BLYP
is likely to apply to many other GGAs.
For example, tests with RPBE, mPWLYP, and BP86 (on the 92 monomers taken from our MD simulation)
all produce one-body energies that are 30-40 meV
smaller than CCSD(T).
Similarly, as with PBE0, the hybrid functionals B3LYP and X3LYP predict rather accurate one-body energies,
coming within 10 meV of CCSD(T).
Of course the conclusion reached here that HF exact exchange is necessary for the proper description
of covalent bond stretching is consistent with what has long been known in the context of
covalent bond breaking (and transition state energies) in the gas
phase (see e.g. Refs. \cite{baker_95,durant_96,lynch_00,Zhao_jpcA_05,Janesko_jcp_08}).
\begin{table}
\caption {\label{tb3} Average differences in the exchange and correlation contributions between the deformed
monomers extracted from liquid water and the gas phase equilibrium monomer structure, obtained with CCSD(T), PBE, and PBE0.
Note that the exchange contribution of CCSD(T) refers to HF exact exchange.
Values in parenthesis are the differences between the two xc functionals and CCSD(T).
Here positive and negative values indicate energy loss and gain, respectively.
All values are in meV.}
\begin{ruledtabular}
\begin{tabular}{cccc}
        & CCSD(T) & PBE & PBE0 \\
\hline
Correlation & --65.5  & +35.2 (--99.7)   & +34.5  (--99.0)  \\
Exchange    & +825.6 & +681.9 (+143.7) & +722.7 (+102.9)\\
\hline
Total       & 761.1  & 717.1 (+44.0) & 757.2 (+3.9) \\
\end{tabular}
\end{ruledtabular}
\end{table}

Several previous studies have examined how DFT xc functionals perform in treating
HBs between water molecules in water clusters
\cite{sms,santra_08,X3LYP_dimer,X3LYP_hexamer_04,joel_pbe,Tsuzuki,Novoa,
xantheas_94,Tschumper_JCPA_06,shields_kirschner,Perdew_JCPB_05,truhlar_pbe1w,truhlar_many-body_06}.
Our study here is somewhat unconventional in that we have examined structures extracted directly from
a liquid water simulation instead of exploring equilibrium gas phase structures.
This has revealed significant differences in how PBE and BLYP perform for the structures
extracted from the liquid compared to the known performance of these functionals for
equilibrium gas phase water clusters.
Thus we see that the behavior of these functionals for e.g. the gas phase equilibrium water dimer
is not a good indicator for how these functionals perform for dimers extracted from the liquid.
One must always exercise caution in making connections between interaction
energies of gas phase clusters and RDFs for the corresponding liquid phase,
particularly in the present circumstances where we have only considered one-- and two--body terms.
Nonetheless, it is plausible that the overbinding observed here for PBE and BLYP, which
originates in $E_{1b}$ errors, is
connected with the overstructuring of these functionals for liquid water.
Indeed, because of the greater error cancelations between the one-- and two--body energies calculated with BLYP,
the overbinding of the dimers from the liquid is less for BLYP than for PBE.
This may again
provide an explanation for why $g_\mathrm{O-O}^{max}$ is less in BLYP compared
to PBE.
This thinking may also explain the low value of $g_\mathrm{O-O}^{max}$ reported in MD simulations with
the RPBE functional \cite{artacho_05,am05_water_09}.
A computed $\sim$50 meV underestimation of the $E_{2b}$ for a gas phase equilibrium water dimer suggests
that the likely errors in $E_{1b}$ for this non-hybrid GGA will be more than compensated for.
In addition, our results are consistent with and help to explain the results from the rigid
water MD simulations \cite{galli_rigid_water,susan_06}.
First, by fixing O-H bonds at or close to the gas phase monomer equilibrium O-H bond length,
the $E_{1b}$ error is eliminated or greatly reduced. Second, the large $E_{2b}$ error
associated with the longest O-H$_d$ bonds is also obviated.

Water molecules in other environments such as those in bulk ice or larger gas phase clusters
will also possess deformed monomers with elongated bonds.
These deformations are smaller than in liquid water but the effect is not negligible.
For example, the average deformation of the monomers in a water hexamer
is 0.05 \AA~ with PBE optimized geometries and in bulk ice Ih PBE predicts
an average deformation of $\sim$0.06 \AA.
Based on the approximate relation between $E_{1b}$ error and deformation established in Fig. \ref{fig3},
such deformations as encountered in small clusters and ice are likely to lead to errors in
$E_{1b}$ of $\sim$30-40 meV.

Finally, we point out that the suggestion that too facile bond stretching may result
in an overstructured liquid is likely to be of relevance to other associated liquids
apart from water.
The relevant experimental and theoretical RDFs of other associated liquids are not as
well established as liquid water.
However, there are indications of BLYP simulations yielding overstructured
RDFs for e.g. liquid ammonia \cite{tuckerman_nh3_99,Dmarx_nh3_03} and
methanol \cite{pagliai_methanol_03,handgraaf_methanol_04}
despite BLYP underestimating the strength of the corresponding gas phase dimers
by $\sim$45\% compared to CCSD(T) \cite{Tsuzuki}. \\

\textbf{V. Summary} \\

In summary, from a PBE simulation of liquid water, monomers and bonded dimers
(from the first coordination shell of the O-O RDF)
were extracted.
With CCSD(T) 75\% of the dimers were shown to be unbound
compared to two gas phase equilibrium water monomers.
This is mainly because the structures of the water monomers inside the liquid differ significantly from
an equilibrium gas phase monomer.
Indeed, with CCSD(T) we find that the average monomer extracted from the PBE
liquid is about 150 meV less stable than
an equilibrium gas phase water monomer.
Among the three xc functionals tested, the two GGAs (BLYP and PBE) underestimate the energy
cost for monomer deformation
(i.e., $E_{1b}$) and
as a consequence BLYP and PBE predict dissociation energies that are too large
by 80 and 43 meV, respectively, compared to CCSD(T).
This is inferior to
the performance of these functionals for the equilibrium water dimer and other water clusters in the gas phase.
Overall PBE0 yields much more accurate dimer dissociation energies,
mainly because it is not susceptible to such large bond stretching errors as the GGAs are.
However, PBE0 is not free from deficiencies in treating the dimers examined here.
Specifically, like the two other functionals, it predicts an increasing error in $E_{2b}$
for the longest O-H$_d$ bonds.
Finally, we have discussed the possible relevance of these results to
DFT simulations of liquid water, to water in other environments, and to other associated liquids.
In particular, we have suggested that the overbinding identified here may provide an explanation
for the overstructured RDFs observed in BLYP and PBE
simulations of liquid water.
However, more work is required to further test this suggestion, with e.g. larger clusters
that give access to higher order terms in the many-body decomposition
and/or clusters embedded in an external electrostatic field
that mimics the remaining water present in the liquid.\\

\textbf{Acknowledgments}\\

This work is supported by the European Commission through the Early
Stage Researcher Training Network MONET, MEST-CT-2005-020908 (\emph{www.sljus.lu.se/monet}).
B.S. is grateful to Martin Fuchs for many helpful discussions.
A.M's work is supported by a EURYI award (\emph{www.esf.org/euryi}),
the EPSRC, and the European Research Council.\\


\begin{thebibliography}{74}
\bibitem[{not({\natexlab{a}})}]{note_PBE_monomer}
\bibinfo{note}{The choice of a PBE optimized monomer as the equilibrium monomer
  reference does not influence the general conclusions arrived at here. For
  example, if a CCSD(T) optimized monomer reference is used, the DFT $E_{1b}$
  errors on average differ by no more than 5 meV from those reported.}
\bibitem[{not({\natexlab{b}})}]{note_blyp_dimers}
\bibinfo{note}{To check if the computed errors in $E_{1b}$ from the xc
  functionals are sensitive to the functional used in the original MD
  simulation a separate 30 ps BLYP MD simulation was performed. From the BLYP
  simulation 12 monomers were extracted and average errors in $E_{1b}$ of 23
  (PBE), 27 (BLYP), and -0.3 meV (PBE0) obtained, compared to average errors in
  $E_{1b}$ of 32 (PBE), 38 (BLYP), and -3 meV (PBE0) as obtained from the PBE
  MD simulation reported.}
\end{thebibliography}

\end{document}